\documentclass[aps,superscriptaddress,showpacs,nofootinbib, 10pt]{revtex4-1}
\usepackage{mathrsfs}
\usepackage{bigints}
\usepackage{latexsym,bm}
\usepackage{graphicx}
\usepackage{indentfirst}
\usepackage{slashed}
\usepackage{amssymb}
\usepackage{amsmath}
\usepackage{bbm}
\usepackage{color}
\usepackage[dvipsnames]{xcolor}
\usepackage{epsfig}
\usepackage[titletoc]{appendix}
\usepackage{multirow}%
\usepackage{rotating}
\usepackage{epstopdf}
\usepackage{extarrows}
\usepackage{tabularx}
\usepackage{soul} 
\usepackage{xcolor}
\usepackage{lipsum}
\usepackage[colorlinks=true,allcolors=BlueViolet,hyperfootnotes=false]{hyperref}

\usepackage[utf8]{inputenc}

\newcommand{\be}{\begin{equation}} \newcommand{\ee}{\end{equation}}
\newcommand{\ba}{\begin{array}{c}} \newcommand{\ea}{\end{array}}
\newcommand{\bea}{\begin{eqnarray}} \newcommand{\eea}{\end{eqnarray}}

\begin{document}
\title{\Large Tau longitudinal and transverse polarizations from visible kinematics in
  (anti-)neutrino nucleus scattering}

\author{E. Hern\'andez}
\affiliation{Departamento de F\'\i sica Fundamental 
  e IUFFyM,\\ Universidad de Salamanca, E-37008 Salamanca, Spain}

\author{J. Nieves}
\affiliation{Instituto de F\'{\i}sica Corpuscular (centro mixto CSIC-UV), Institutos de Investigaci\'on de Paterna,
Apartado 22085, 46071, Valencia, Spain}

\author{F. S\'{a}nchez} \affiliation{Universit\'{e} de Gen\`{e}ve -
  Facult\'{e} des Sciences, D\'{e}partement de Physique Nucl\'{e}aire
  et Corpusculaire (DPNC) \\ 24, Quai Ernest-Ansermet, CH-1211
  Gen\`{e}ve 4, Switzerland}

\author{J. E. Sobczyk}
\affiliation{Institut f\"ur Kernphysik and PRISMA+ Cluster of 
Excellence, Johannes Gutenberg-Universit\"at Mainz, 55128 Mainz, Germany}

\date{\today}

\begin{abstract}
 
Since the $\nu_\tau(\bar\nu_\tau) A_Z \to \tau^\mp X$ reaction is notoriously difficult to be directly measured, the information on the dynamics of this nuclear process  should be extracted from the analysis of the energy and angular distributions of the tau decay visible products. These distributions depend, in addition to  $d^{\,2}\sigma/(dE_\tau d\cos\theta_\tau)$,  on  the components of the tau-polarization vector. 
We give, for the first time, the general expression for the outgoing hadron (pion or rho meson) energy and angular differential cross section for the sequential $\nu_\tau A_Z \to \tau^-(\pi^- \nu_\tau, \rho^-\nu_\tau) X$ and $\bar\nu_\tau A_Z \to \tau^+(\pi^+ \bar\nu_\tau, \rho^+ \bar\nu_\tau) X$ reactions. 
Though all possible nuclear reaction mechanisms contribute to the distribution, it may be possible to isolate/enhance one of them by implementing appropriate selection criteria.
For the case of the quasi-elastic reaction off oxygen and neutrino energies below 6 GeV, we show that the pion distributions are quite sensitive to the details of the  tau-polarization components. We find significant differences  between the full calculation, where the longitudinal and transverse  components of the tau polarization vector vary with the energy and the scattering angle of the produced tau, and the simplified scheme in which the polarizations are  set to one and zero, being the latter their respective asymptotic values reached in the high energy regime.  In addition to its potential impact on neutrino oscillation analyses, this result  can   be used to further test different nuclear models, since these  observables provide complementary information to that obtained by means of the inclusive nuclear weak charged-current differential cross section. We also study the effects on the cross section of   the  $W_4$ and $W_5$ nuclear structure functions, which contributions are proportional to the charged lepton mass, and therefore difficult to constrain in muon and electron neutrino experiments. 
\end{abstract}

\maketitle
\section{Introduction}

The outgoing $\tau$-lepton produced  in $\nu_\tau$ and $\bar\nu_\tau$ charged-current (CC) nuclear interactions is not fully polarized for a wide range of energies. Within the Standard Model (SM), the tau polarization component perpendicular to the lepton scattering plane is zero 
\cite{LlewellynSmith:1971uhs,Kuzmin:2004yb}, while the $\tau^\mp$  longitudinal and transverse  polarizations within this plane do not vanish, being
 sensitive to independent combinations of  nuclear structure functions~\cite{Hagiwara:2003di, Valverde:2006yi}. Thus, these observables can be used to further test different nuclear models, since they provide complementary information to that obtained by means of the inclusive nuclear CC differential cross section. 

Longitudinal and transverse tau polarization projections have been previously computed in the quasi-elastic (QE) region, in which the single nucleon knock-out is the dominant reaction mechanism. The pioneering work of Ref.~\cite{Hagiwara:2003di} considered the nucleus as an ensemble of free nucleons, and predictions were substantially improved in Ref.~\cite{Graczyk:2004uy}, with the inclusion of Random Phase Approximation (RPA) and effective nucleon mass effects. More robust theoretical results were presented in \cite{Valverde:2006yi}, and in particular in \cite{Sobczyk:2019urm},  for neutrino energies below 10 GeV, by using realistic spectral functions\footnote{The use of an effective mass for the nucleon is a simplified method to account for the effects due to the change of its dispersion relation inside the nuclear medium. A proper description, however, is achieved by dressing the nucleon propagators and constructing realistic particle and hole spectral functions, which incorporate dynamical effects  that depend on both the energy and momentum of the 
nucleons~\cite{Nieves:2017lij}.}.  Recently, the effects on the $\nu_\tau +n \to \tau^- p$  and $\bar\nu_\tau p \to \tau^+ n$ CCQE scattering of  different
parametrizations of the isovector vector, axial-vector and  pseudoscalar form factors together with  the use of second class currents with time-reversal invariance have been analyzed in Ref.~\cite{Fatima:2020pvv}.  The cross section for $\nu_\tau/\bar\nu_\tau$ scattering  off nuclei and the polarization state of the outgoing $\tau^\mp$ were also studied in \cite{Hagiwara:2003di} for $\Delta$ resonance production and deep inelastic scattering (DIS)
processes, within the simplified picture for the nucleus employed in that work. A more rigorous treatment of the nuclear medium effects  in the DIS region is presented in Ref.~\cite{Zaidi:2021iam}, where the authors have obtained results for the scattering cross sections in  $^{40}$Ar in the energy region of interest for the proposed DUNE experiment~\cite{Machado:2020yxl}.

The $\tau$ lepton decays rapidly (with a mean life of $2.9\times 10^{-13}$\,s) and its disintegration involves  at least one  neutrino that escapes detection. This makes   
its clear identification very challenging   posing a serious problem  to the experimental measurement  
of the inclusive nuclear $\nu_\tau(\bar\nu_\tau) A_Z \to \tau^\mp X$  differential cross section $d^{\,2}\sigma/(dE_\tau d\cos\theta_\tau)$.  Moreover, the polarization state  of the outgoing tau cannot be directly measured.

\begin{figure*}[b]
\centering
\includegraphics[scale=0.35]{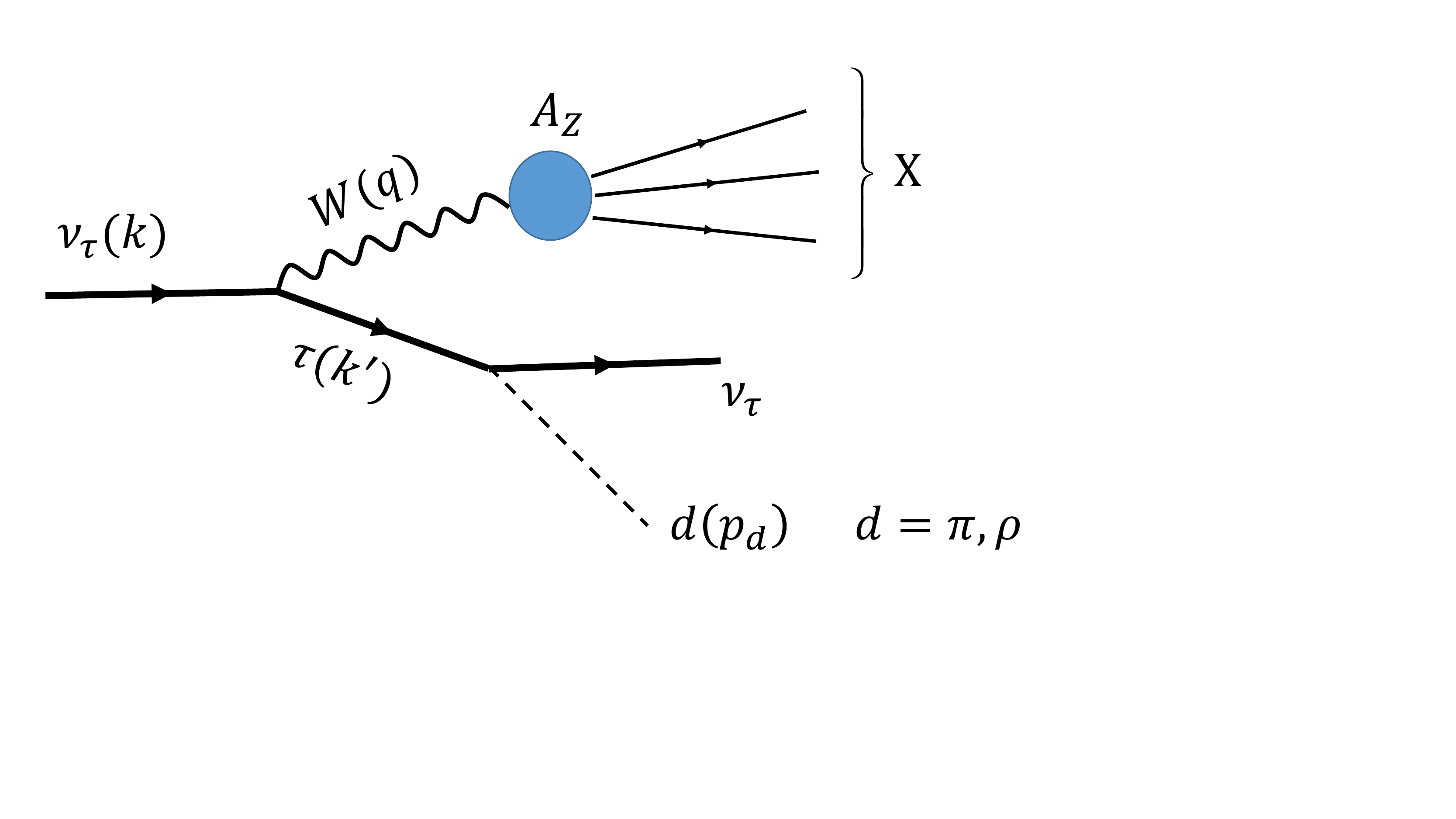}
\vspace{-2cm}
\caption{Sequential $\nu_\tau A_Z \to \tau^-(\pi^- \nu_\tau,\rho^-\nu_\tau) X$ nuclear reaction.}
\label{fig:diagram}
\end{figure*}

The information about the  $\tau$ lepton polarization should be inferred from the energy and angular distributions of its decay visible products. The study of these distributions and their relation to the components of the tau-polarization vector and to $d^{\,2}\sigma/(dE_\tau d\cos\theta_\tau)$ is precisely the main objective of this work. We give, for the first time, the general expression for the outgoing hadron (pion or rho meson) energy and angular differential cross section for the sequential $\nu_\tau A_Z \to \tau^-(\pi^- \nu_\tau,
  \rho^-\nu_\tau) X$ and $\bar\nu_\tau A_Z \to \tau^+(\pi^+
  \bar\nu_\tau, \rho^+ \bar\nu_\tau) X$ reactions (see Fig.\ref{fig:diagram}).
We show that such a distribution, given in Eq.~\eqref{eq:defsigmad}, depends on the tau inclusive nuclear CC differential cross section and the longitudinal and transverse polarization observables integrated, with certain dynamical weights, over the  outgoing $\tau$ available phase space. All possible nuclear reaction mechanisms contribute to the $d^2\sigma_d/(dE_d d\cos\theta_d)$ visible distribution (where $d=\pi,\, \rho$). However, depending on the neutrino energy and implementing appropriate event selections, it may be possible to isolate/enhance the contribution of different (anti-)neutrino-nucleus reaction channels (QE, 2p2h, coherent and incoherent pion production, DIS, etc.) within some regions of the visible $(E_d,\cos\theta_d)$ phase-space. 
  
Since in the  $E_\nu \gg m_\tau$ high energy limit the outgoing $\tau$ leptons are 
produced in fully polarized states, the interesting energy region to learn details on the 
polarization observables is limited to the values of $E_\nu \lesssim 10$ GeV. This  energy 
range can be studied by the DUNE oscillation experiment~\cite{Machado:2020yxl}, although the 
measurement is demanding because of  low statistics and the contamination of the sample
by neutrino's neutral-current (NC) interactions. 
For these relatively moderate energies of interest, we want to illustrate the visible distributions that can be obtained and to emphasize on the relevance of  a  non-trivial tau-polarization vector. To this end we show results  for the pion mode in oxygen, assuming a pure QE-reaction mechanism evaluated within the model derived in Ref.~\cite{Nieves:2004wx}. 
We find significant differences in the pion distributions between the full calculation, where the tau longitudinal and transverse polarization components depend on the nuclear model, and the simplified scheme in which they are set to  one and zero (their respective asymptotic values reached in the high energy regime). 
Even if the direct determination of these distributions might be beyond the capabilities of next generation experiments, having their best theoretical values could be very relevant in the analysis of certain experiments. For instance, the expected sizeable contribution of low energy $\nu_{\tau}$'s in the oscillated samples of the DUNE experiment~\cite{Machado:2020yxl} might affect the precision of the oscillation parameters since the momentum and angular distributions of the pions, but also of the  muons and electrons from the tau leptonic decays, will influence the ability of the experiment to control the backgrounds at the far detector. 

This work is organized as follows. In Sec.~\ref{sec:unpol} we  discuss the inclusive $\nu_\tau(\bar\nu_\tau) A_Z \to \tau^{\mp} X$  cross sections in terms of the structure functions, which provide a general parametrization of the nuclear hadron tensor. We also introduce the $\tau^{\mp}$ polarization vector and give the expressions of their components as a function of the structure functions. 
In Sec.~\ref{sec:sequent}, we present the master formula for the cross section of the 
$\nu_\tau A_Z \to \tau^-(\pi^- \nu_\tau,
  \rho^-\nu_\tau) X$ and $\bar\nu_\tau A_Z \to \tau^+(\pi^+
  \bar\nu_\tau, \rho^+ \bar\nu_\tau) X$ sequential processes.
  Results for the CCQE contribution evaluated in oxygen at different (anti-)neutrino energies are presented in Sec.~\ref{sec:results} and a brief summary of the main findings of our work is given in Sec.~\ref{sec:summary}.

\section{Unpolarized and polarized nuclear inclusive cross section}
\label{sec:unpol}
We will first study the CC nuclear inclusive reactions, 
\begin{equation}
\nu_\tau A_Z \to \tau^- X, \quad \bar\nu_\tau A_Z \to \tau^+ X
\end{equation}
where a  tau (anti-)neutrino, 
with four momentum $k^\mu=(E_\nu,\vec{k})$,  exchanges a $W$ boson with an atomic
nucleus with initial momentum $P^\mu=(M_A,\vec{0}\,)$, and a 
lepton $\tau^-$ (or $\tau^+$) is detected with four-momentum
$k'{}^\mu=(E_\tau,\vec{k'})$. In these processes, the final hadronic
state is not detected, and the unpolarized differential cross section
in the laboratory frame 
reads \cite{Nieves:2004wx}
\begin{equation} 
\label{eq:sigma}
\Sigma_0^{(\nu_\tau, \bar \nu_\tau)} \equiv \frac{d^{\,2}\sigma_{(\nu_\tau, \bar \nu_\tau)}}{dE_\tau d\cos\theta_\tau} =
\frac{|\vec{k}^\prime|G^2_F M_A}{\pi} F_{(\nu_\tau, \bar \nu_\tau)},
\end{equation}
where $G_F$ is the Fermi weak coupling constant, $\cos\theta_\tau \in
\, ]-1,1]$ is the angle between $\vec{k}$ and $\vec{k'}$, $m_\tau \le E_\tau \le E_\nu $ \footnote{We are not
      explicitly considering the minimum energy ($\sim$ tens of MeV) to be transferred to the
    nuclear system to account for the mass difference between the mass of
    the initial nucleus and that of the ground state of the final
    nuclear configuration~\cite{Nieves:2004wx, Bourguille:2020bvw}. This should be a good approximation for
    $E_\nu$, at least, in the few GeV region.  } and $F$ is defined as 
\begin{eqnarray}
F_{(\nu_\tau, \bar \nu_\tau)} & = & 
\left( 2W_1 +\frac{m_\tau^2}{M_A^2}W_4\right)
(E_\tau-|\vec{k}^\prime|\cos\theta_\tau)
+ W_2(E_\tau+|\vec{k}^\prime|\cos \theta_\tau) 
\nonumber\\
&&\mbox{}
- W_5\frac{m_\tau^2}{M_A} \mp \frac{W_3}{M_A}
\left(  E_{\nu}E_\tau+|\vec{k}^\prime|^2
       -(E_{\nu}+E_\tau)|\vec{k}^\prime|\cos\theta_\tau
\right)
\label{eq:defF}
\end{eqnarray}
where the $\mp$ sign in the $W_3$ term  correspond to the case of neutrino or anti-neutrino scattering. 
The real Lorentz-scalar structure functions $W_i(q^0, q^2)$, depend on the four-momentum
transferred to the nuclear system, with $q^2=(q^0)^2-|\vec{q}\,|^2$, $q^0=(E_\nu-E_\tau)$ and
$|\vec{q}\,|=\left(E_\nu^2+ \vec{k}^{\prime
  2}-2E_\nu|\vec{k}'|\cos\theta_\tau\right)^\frac12$, and they are obtained from the decomposition
of the hadronic tensor~\cite{Nieves:2004wx}
\begin{equation}
\frac{W^{\mu\nu}}{2M_A} = - g^{\mu\nu}W_1 + \frac{P^\mu
  P^\nu}{M_A^2} W_2 + {\rm i}
  \frac{\epsilon^{\mu\nu\gamma\delta}P_\gamma q_\delta}{2M_A^2}W_3 +  
\frac{q^\mu  q^\nu}{M_A^2} W_4 + \frac{P^\mu q^\nu + P^\nu q^\mu}
{2M_A^2} W_5 + i\frac{P^\mu q^\nu - P^\nu q^\mu}
{2M_A^2} W_6
\label{eq:hadron_general}
\end{equation}
with $\epsilon_{0123}= +1$ and the metric $g^{\mu\nu}=(+,-,-,-)$.
The structure functions are different for neutrino or anti-neutrino reactions because the role of protons and neutrons is exchanged. However, for simplicity in
the notation, they are shown without the  $(\nu_\tau, \bar \nu_\tau)$
label.  
 The
term proportional to $W_6$ does not contribute 
to the double differential cross section and, thus, it does not appear
in the full expression for $F$ in Eq.~\eqref{eq:defF}. As mentioned,
we follow here the formalism and conventions of
Ref.~\cite{Nieves:2004wx}.\footnote{We should mention that there is a typo
in Eq.~(10) of this latter work, which affects  the $W_4$ term, where $\sin^2\theta_\tau$ should be $\sin^2\theta_\tau/2$.}

The  (anti)neutrino inclusive-differential cross section  for the production of an (anti-)tau with polarization  $h=\pm1$ along a certain four-vector $S^\mu$, verifying $S^2=-1$ and $S\cdot k'=0$~\cite{Penalva:2021gef}, can be 
written, using the results of Appendix A of
Ref.~\cite{Sobczyk:2019urm}, as
\begin{equation}
\Sigma^{\nu_\tau} = \frac12\Sigma_0^{\nu_\tau}\left(1+ h S_\mu {\cal
  P}^\mu_{(\tau)}\right), \quad \Sigma^{\bar\nu_\tau} =
\frac12\Sigma_0^{\bar\nu_\tau}\left(1- h S_\mu {\cal
  P}^\mu_{(\bar\tau)}\right)
\label{eq:defsecpol}
\end{equation}
where $\Sigma_0^{(\nu_\tau\bar, \nu_\tau)}$, given in
Eq.~\eqref{eq:sigma}, is the unpolarized cross section, and ${\cal P}^\mu_{(\tau, \bar\tau)}$ is the
(anti-)tau lepton polarization
vector~\cite{Penalva:2021gef,Penalva:2021wye}, which is given  in
Eq.~(5) of Ref.~\cite{Sobczyk:2019urm} for SM neutrino and anti-neutrino nuclear
inclusive reactions. Note that the anti-neutrino $ {\cal
  P}^\mu_{(\bar\tau)}$ vector introduced here in Eq.~\eqref{eq:defsecpol} above differs by a minus sign from that defined in Ref.~\cite{Sobczyk:2019urm}. In the absence of physics
beyond the SM,  the relevant
components of the polarization vector in the laboratory system are
denoted as $P_L$ (longitudinal, in the direction of
$\vec{k'}$), and $P_T$ (transverse to $\vec{k'}$ and contained in the
neutrino-tau lepton plane),  
\begin{equation}
P_{L,T}^{(\tau, \bar\tau)} = - \left( {\cal P}_{(\tau,
  \bar\tau)}\cdot  N_{L,T}\right), \quad N^\mu_L =
\left(\frac{|\vec{k}'|}{m_\tau},
\frac{E_\tau\vec{k}'}{m_\tau|\vec{k}'|} \right), \quad N^\mu_T =
\left(0, \frac{(\vec{k}\times\vec{k}')\times
  \vec{k}'}{|(\vec{k}\times\vec{k}')\times \vec{k}'|}\right) \label{eq:defPLPT}
\end{equation}
The $P_{L,T}$ components depend on the lepton kinematics and on the
(anti-)neutrino structure
functions, $W_i$, introduced in
Eq.~(\ref{eq:hadron_general})~\cite{Valverde:2006yi,Hagiwara:2003di,Sobczyk:2019urm}.
 For the sake of completeness and clarity, we also
reproduce these formulae here\footnote{Note that the projections used in  Eq.~\eqref{eq:defPLPT} to define $P_{L,T}^{(\tau, \bar\tau)}$ 
  differ, for both neutrino (tau) and
  anti-neutrino (anti-tau) reactions, in
  a global sign to those taken in Refs.~\cite{Valverde:2006yi,Sobczyk:2019urm} (see Eq.~(6) of Ref.~\cite{Sobczyk:2019urm}). In addition, the
  convention employed here for the polarized cross section in
  Eq.~\eqref{eq:defsecpol}, changes the sign of the  anti-tau
  polarization components $P_{L,T}^{\bar\tau}$ with respect to those
  given in Refs.~\cite{Valverde:2006yi,Sobczyk:2019urm}. In summary, tau $P_{L,T}^{\tau}$ [anti-tau $P_{L,T}^{\bar\tau}$] defined here differ in  sign [are equal] to the ones introduced in Refs.~\cite{Valverde:2006yi,Sobczyk:2019urm}, as can be seen comparing Eqs.~\eqref{eq:PL} and \eqref{eq:PT} below with Eqs.~(5) and (6) of Ref.~\cite{Valverde:2006yi}.},
\begin{eqnarray}
P_L^{(\tau, \bar\tau)} &=& 
 \left\{ 
\left(2W_1 - \frac{m_\tau^2}{M_A^2}W_4\right)
  (|\vec{k}^\prime|-E_\tau\cos\theta_\tau)
    + W_2(|\vec{k}^\prime|+E_\tau\cos\theta_\tau)     
    - W_5\frac{m_\tau^2}{M_A}\cos\theta_\tau  
\right. \label{eq:PL}\\ \nonumber
& & \left. 
    \mp \frac{W_3}{M_A}((E_\nu+E_\tau)|\vec{k}^\prime|
           -(E_\nu E_\tau+|\vec{k}^\prime|^2)\cos\theta_\tau) 
\right\} /F_{(\nu_\tau, \bar\nu_\tau)} \\
P_T^{(\tau, \bar\tau)} &=&
  m_\tau\sin\theta_\tau
   \left(2W_1 - W_2 - \frac{m_\tau^2}{M_A^2}W_4 + W_5\frac{E_\tau}{M_A}
    \mp W_3\frac{E_\nu}{M_A}\right)/F_{(\nu_\tau, \bar\nu_\tau)} \label{eq:PT}
\end{eqnarray}
where the $\mp$ sign in the $W_3$ term  correspond to the case of tau  or
anti-tau polarization components. Note that the transverse polarization $P_T$ is
proportional to the charged-lepton mass, and therefore for muon or
electron neutrino reactions, it is highly suppressed, as expected by
conservation of chirality~\cite{Sobczyk:2019urm}.

The study of the tau polarization vector in $(\nu_\tau,\tau)$ and
$(\bar\nu_\tau,\bar\tau)$ reactions is of great theoretical interest,
since both $P_L$ and $P_T$ display peculiar sensitivities to the ingredients of the nuclear and reaction
models, which are different to the ones shown by the pure double differential inclusive cross sections. This is to
say, the polarization observables of Eqs.~\eqref{eq:PL} and
\eqref{eq:PT}, and the unpolarized function $F$ defined in Eq.~\eqref{eq:defF} are given by different linear
independent combinations of the hadron-tensor $W_{1,2,3,4,5}$
structure functions, being sensitive to different
$(E_\tau,\cos\theta_\tau)$ kinematics. Experimental information on $P_{L,T}^{(\tau, \bar\tau)}$  would provide additional valuable constraints to test the models used to describe (anti-)neutrino-nucleus cross sections.

\section{Sequential $\nu_\tau A_Z \to \tau^-(\pi^- \nu_\tau,
  \rho^-\nu_\tau) X$ and $\bar\nu_\tau A_Z \to \tau^+(\pi^+
  \bar\nu_\tau, \rho^+ \bar\nu_\tau) X$ reactions}
  \label{sec:sequent}
  
Given the practical impossibility of   making a direct measurement of the $\tau$ polarization, the information on the polarization state of the outgoing tau-lepton, encoded in its polarization vector, should be  obtained from the energy and angular distributions of its visible decay products.  A more sensitive measurement would be obtained from a multiple differential cross section involving also the tau-momentum variables. However, such a measurement will certainly suffer from  smaller statistics, besides the fact that fully reconstructing the final  $\tau$ momentum  represents an experimental challenge, because it might not travel far enough for a displaced vertex and its decay
involves at least one  neutrino. 

However, even if the three momentum of tau could not be measured, information about the  $\nu_\tau (\bar\nu_\tau) A_Z \to \tau^\mp X$ inclusive nuclear processes and on the polarization state in which the tau-lepton is created can be extracted from
the distribution of its charged decay products. This is also the
case for the semileptonic decays of bottomed hadrons into charmed ones,
driven by the $b \to  c \tau\bar\nu_\tau$ transition (see for instance
Refs.~\cite{Penalva:2021wye} and \cite{Penalva:2022vxy}, in which   this section is based). The three
dominant $\tau$ decay modes $\tau \to \pi \nu_\tau ,\, \rho \nu_\tau, \,
\ell\bar\nu_\ell\nu_\tau$ ($\ell=e,\mu$) account for more than 70\% of the total
$\tau$ decay width ($ \Gamma_\tau$).  In this section, we will pay
attention to the hadron-modes of the tau decay, and study the
visible energy and angular distributions of the pion or rho mesons in the
nuclear inclusive processes
\begin{eqnarray}
&&\nu_\tau A_Z \to  \tau^- X \hspace{3cm} \bar\nu_\tau A_Z \to  \tau^+ X  \nonumber \\
&&\hspace{1.5cm} \searrow   \nu_\tau \pi^-, \,\, \nu_\tau \rho^- \hspace{2.1cm}
\, \searrow \,  \bar\nu_\tau \pi^+, \,\, \bar\nu_\tau \rho^+
\end{eqnarray}
The analog reactions induced by the leptonic tau-decay channel are less sensitive to  the tau-polarization components, with their contribution being around  a factor of 3 smaller than for the $\tau\to \pi \nu_\tau$ case~\cite{Penalva:2022vxy}. Hence, the $\tau \to \ell\bar\nu_\ell\nu_\tau$ decay will provide less statistically meaningful results.

Following Sec.~3 of  Ref.~\cite{Penalva:2021wye},  we find that for any of these hadron
modes, $(\nu_\tau, \bar \nu_\tau)A_Z \to X \tau^\mp  \to X
d^\mp(\nu_\tau, \bar \nu_\tau) $ [$d=\pi,\rho$],  the laboratory double differential cross
section with respect to the outgoing hadron energy ($E_d$) and the
cosinus of the angle ($\theta_d$)
formed by the hadron ($\vec{p}_d$) and the incoming (anti-)neutrino ($\vec{k}$) three-momenta, reads (see Fig.~\ref{fig:diagram})
\begin{eqnarray}
\frac{d^2\sigma_d^{(\nu_\tau, \bar
  \nu_\tau)}}{dE_d d\cos\theta_d} & = &
     {\cal B}_d
     \frac{m^2_\tau}{m^2_\tau-m^2_d}\frac{G^2_F M_A}{\pi^2}\int_{E_\tau^-}^{E_\tau^{\rm
     sup}} dE_\tau \int_{\cos(\theta_d+\theta_{\tau
         d})}^{\cos(\theta_d-\theta_{\tau d})}
     \frac{d(\cos\theta_\tau)F_{(\nu_\tau, \bar
  \nu_\tau)}(E_\tau,\cos\theta_\tau)}{\sqrt{\left[\cos(\theta_d-\theta_{\tau
             d})-\cos\theta_\tau\right]\left[\cos\theta_\tau-\cos(\theta_d+\theta_{\tau
             d})\right]}}\nonumber  \\
\nonumber  \\
& \times& \left \{1+ \frac{2m_\tau}{m^2_\tau-m^2_d}
a_d\left[P^{(\tau,
    \bar\tau)}_L(E_\tau,\cos\theta_\tau)\left(\frac{E_d|\vec{k}'|}{m_\tau}-\frac{E_\tau|\vec{p}_d|}{m_\tau}\cos\theta_{\tau
             d}
  \right)\right.\right.\nonumber  \\
&&
\left.\left.+\frac{P^{(\tau,
    \bar\tau)}_T(E_\tau,\cos\theta_\tau)}{\sin\theta_\tau}|\vec{p}_d|\left(\cos\theta_d-\cos\theta_\tau
\cos\theta_{\tau d} \right)\right] \right\} \label{eq:defsigmad}
\end{eqnarray}
In the above expression, ${\cal B}_{d=\pi,\rho}$ is the branching fraction for the $\tau \to
\pi\nu_\tau$ or $\tau \to \rho\nu_\tau$ decays, $m_d=m_\pi$ or $m_\rho$,
$a_{d=\pi}=1$ or
$a_{d=\rho}=(m_\tau^2-2m_\rho^2)/(m_\tau^2+2m_\rho^2)$, and
$\theta_{\tau d}$ is the angle formed by the $\tau$ and outgoing
hadron three momenta,
\begin{equation}
\cos\theta_{\tau d}= \frac{2E_\tau E_d-m^2_\tau -m^2_d}{2|\vec{k}'||\vec{p}_d|}\,.
\end{equation}
In addition, we introduce
\begin{equation}
 E_\tau^\pm = \frac{(m^2_\tau+m_d^2)E_d \pm (m^2_\tau-m_d^2)|\vec{p}_d|}{2m_d^2}, \quad E_d^{\rm max,\, int}= \frac{(m^2_\tau+m_d^2)E_\nu \pm (m^2_\tau-m_d^2)\sqrt{E_\nu^2-m_\tau^2}}{2m_\tau^2}
\end{equation}
where the $\pm$ in $E_d^{\rm max,\, int}$ correspond to the  maximum energy reachable by the $d-$hadron (max) and to some intermediate one (int), which will be relevant in the discussion below, respectively. The minimum value of $E_d^{\rm int}$ turns out to be $m_d$ for $E_\nu=(m^2_\tau+m_d^2)/(2m_d)$. The lower limit of the $E_\tau$ integration is always $E_\tau^-$, which is  bigger or equal than the tau mass for any value of $E_d$. The upper limit could be either $E_\nu$ or $E_\tau^+$. Actually, we have 
\begin{eqnarray}
 E_\nu \le \frac{m^2_\tau+m_d^2}{2m_d} & \Rightarrow & E_d^{\rm int}\le E_d \le E_d^{\rm max}, \quad E_\tau^{\rm
     sup} = E_\nu \nonumber \\ \nonumber \\
     E_\nu > \frac{m^2_\tau+m_d^2}{2m_d} & \Rightarrow & m_d\le E_d \le E_d^{\rm max}, \quad E_\tau^{\rm
     sup} = H(E_d^{\rm int}-E_d)E_\tau^+ +  H(E_d-E_d^{\rm int})E_\nu
\end{eqnarray}
with $H(...)$, the step function. Finally, there is no limitation for the outgoing hadron angle and  $\cos\theta_d \in
\, ]-1,1]$.

In Eq.~\eqref{eq:defsigmad}, it is assumed that the $\tau$ lepton exits the nucleus before decaying, and thus the outgoing pion or rho meson does not suffer any further final state interaction.

As shown in Ref.~\cite{Penalva:2021wye}, the contribution in Eq.~\eqref{eq:defsigmad} which depends on the polarization components vanishes after integrating over the energies and angles of the visible hadron, while the independent one is properly normalized such that
\begin{equation}
 \int dE_d \, d\cos\theta_d\frac{d^2\sigma_d^{(\nu_\tau, \bar
  \nu_\tau)}}{dE_d d\cos\theta_d}= {\cal B}_d \int_{m_\tau}^{E_\nu} dE_\tau \int_{-1}^{+1}d\cos\theta_\tau \frac{d^{\,2}\sigma_{(\nu_\tau, \bar \nu_\tau)}}{dE_\tau d\cos\theta_\tau} \equiv {\cal B}_d \,\sigma_{(\nu_\tau, \bar \nu_\tau)}(E_\nu), \quad  E_\nu \ge m_\tau \label{eq:norm}
\end{equation}
with $\sigma_{(\nu_\tau, \bar \nu_\tau)}(E_\nu)$ the unpolarized tau (anti-)neutrino nuclear inclusive cross section
in the laboratory frame obtained from the integration of the differential distribution of Eq.~\eqref{eq:sigma}.

Note that the maximal information that one can in principle extract would be the four dimensional differential cross section $d^4\sigma_d^{(\nu_\tau, \bar
  \nu_\tau)}/(dE_d d\cos\theta_d dE_\tau d\cos\theta_\tau)$, which would require the measurement of the tau momentum. 
  Such unfolded distribution might allow to obtain all three $F_{(\nu_\tau, \bar
  \nu_\tau)}(E_\tau,\cos\theta_\tau), P_L^{(\tau, \bar\tau)}(E_\tau,\cos\theta_\tau)$ and 
  $P_T^{(\tau, \bar\tau)}(E_\tau,\cos\theta_\tau)$ observables, which otherwise appear in 
  Eq.~\eqref{eq:defsigmad} within an integral over the tau kinematical variables.

The expression of Eq.~\eqref{eq:defsigmad} for the energy and angular distribution of the 
hadron product after the decay of the  virtual $\tau$ is general, and it implies a sum  
over all possible nuclear reaction mechanisms. Nevertheless, depending on the neutrino 
energy and implementing some cuts, it may be possible to isolate/enhance the contribution 
to the inclusive $d^2\sigma_d^{(\nu_\tau, \bar
  \nu_\tau)}/(dE_d d\cos\theta_d)$ cross section of different (anti-)neutrino-nucleus 
  reaction channels  (QE, 2p2h, coherent and incoherent pion production, DIS, etc.) for 
  some regions of the visible $(E_d,\cos\theta_d)$ phase-space. 
  
  The detailed study of  $d^2\sigma_d^{(\nu_\tau, \bar  \nu_\tau)}/(dE_d d\cos\theta_d)$  
  provides some additional  observables, which will serve to further test the nuclear and 
  reaction models used to describe the CC (anti-)neutrino-nucleus interactions. Moreover 
  and as mentioned above, the $(E_d,\cos\theta_d)$ distribution should be much easier to 
  measure experimentally than the polarized $\nu_\tau A_Z \to \tau^- X$ and 
  $\bar\nu_\tau A_Z \to \tau^+ X$ differential cross sections. Note, however, that there 
  will be also pions or rho-mesons produced in the nuclear part of the interaction, 
  including the re-scattering of the produced nucleons, pions, etc. during their path 
  exiting the nuclear environment. One should look at the $(E_d,\cos\theta_d)-$pattern 
  provided by them, and focus the study in kinematical regions for which the number of 
  nuclear events is much smaller than that due to  pions or rho-mesons coming from the tau 
  decay. 
  
  Another source of unwanted background is  the high contamination of muon and/or electron 
  (anti-)neutrinos that will be likely  present in the available $\nu_\tau$ and 
  $\bar \nu_\tau$ beams. The nuclear $\nu_{\mu,e}$ and $\bar \nu_{\mu,e}$ interactions can 
  give rise to an abundant production of pions and rho mesons. Though CC processes  will 
  produce an outgoing muon or electron which can be used to discard the event  through 
  transverse momentum balance, NC reactions represent a more challenging background to 
  the  $d^2\sigma_d^{(\nu_\tau, \bar  \nu_\tau)}/(dE_d d\cos\theta_d)$ distribution, since 
  the outgoing neutrino will not be detected. Given an experimental setup, these events, 
  not coming from the $\tau$ decay,  need to be taken into account for a meaningful analysis.
  
 Note that this muon/electron neutrino NC background should no affect sequential decays 
 initiated by an anti-neutrino $\bar\nu_\tau$
scattered off a bound-nucleon producing in the final state a $\tau^+$,  together with a 
hyperon $Y=\Lambda, \Sigma$. Though Cabibbo suppressed, the reaction 
$\bar\nu_\tau A_Z \to \tau^+(\pi^+ \bar\nu_\tau, \rho^+ \bar\nu_\tau) Y+ X$
with the detection of the hyperon,  and no kaons in the final state, should be considerably 
less affected by this unwanted NC background. However, in this case, the nuclear structure 
functions will  be different to those probed in the Cabibbo allowed transitions, beginning  
from the fact that the outgoing hyperon will not be Pauli blocked~\cite{Singh:2006xp, Sobczyk:2019uej}.  

In experimental analyses, it is common to use the longitudinal ($p_{Ld}$) and transverse  
($p_{Td}$) pion/rho momentum components which are related to $E_d$ and $\theta_d$ through 
the relations
\begin{eqnarray}
 p_{Ld} = \sqrt{E_d^2-m_d^2}\,\cos\theta_d, \qquad p_{Td} = \sqrt{E_d^2-m_d^2}\,\sin\theta_d 
 \end{eqnarray}
with  $E_d^2= (m_d^2+p_{Ld}^2+p_{Td}^2)$. From the Jacobian of the transformation, we find 
for the corresponding double-differential cross section
\begin{equation}
\frac{d^2\sigma_d^{(\nu_\tau, \bar
  \nu_\tau)}}{dp_{Ld}\,dp_{Td}} = \frac{p_{Td}}{(p_{Ld}^2+p_{Td}^2)^\frac12}\frac{1}{(p_{Ld}^2+p_{Td}^2+m_d^2)^\frac12} \left[\frac{d^2\sigma_d^{(\nu_\tau, \bar
  \nu_\tau)}}{dE_d d\cos\theta_d}\right]= \frac{\sin\theta_d}{E_d} \left[\frac{d^2\sigma_d^{(\nu_\tau, \bar
  \nu_\tau)}}{dE_d d\cos\theta_d}\right]
  \label{eq:dsdpldpt}
\end{equation}  
\section{Results}\vspace{.15cm}
\label{sec:results}

\begin{figure*}
\begin{center}
\hspace*{-2cm}\includegraphics[scale=0.4]{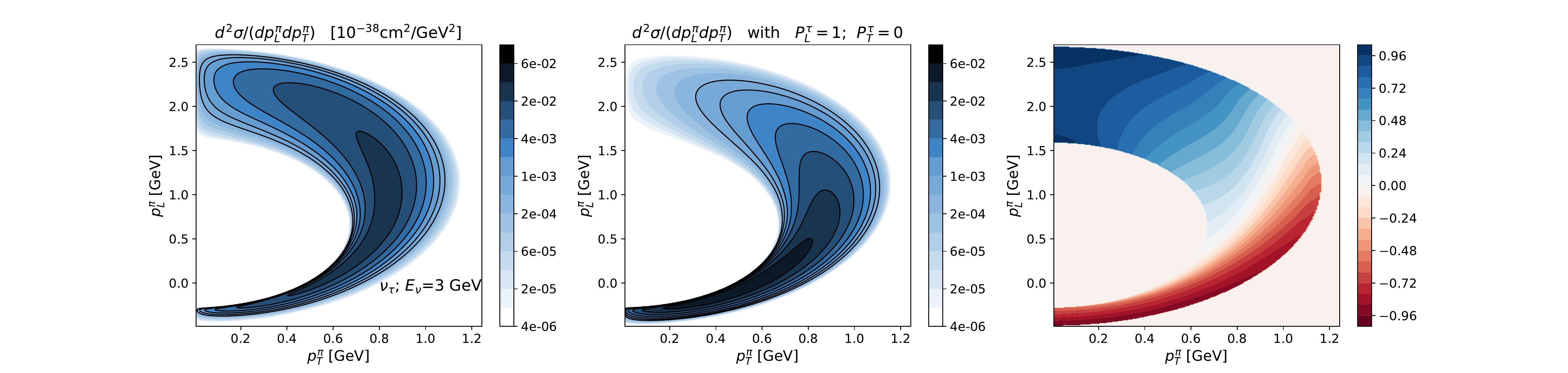}\\
\hspace*{-2cm}\includegraphics[scale=0.4]{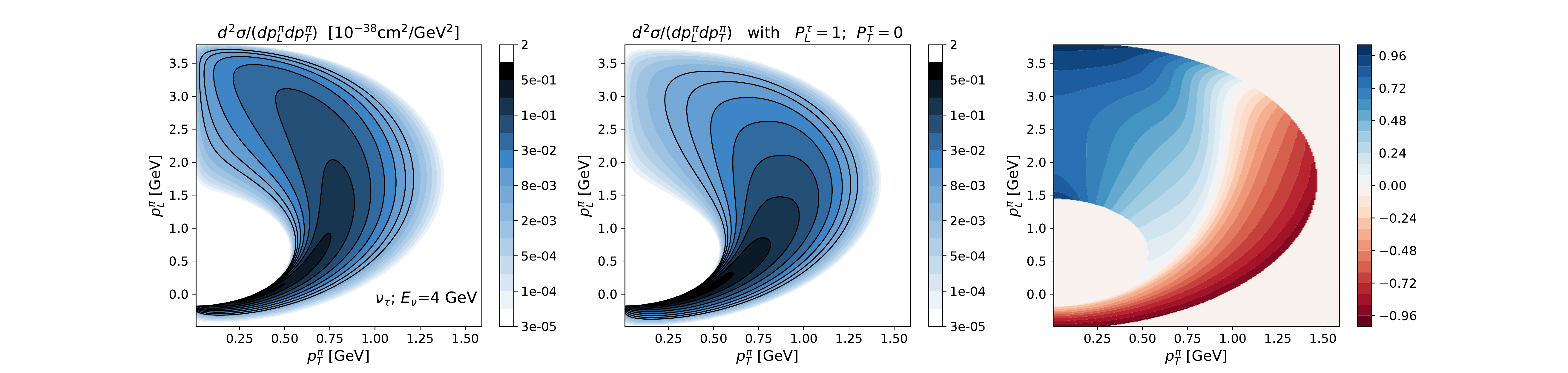}\\
\hspace*{-2cm}\includegraphics[scale=0.4]{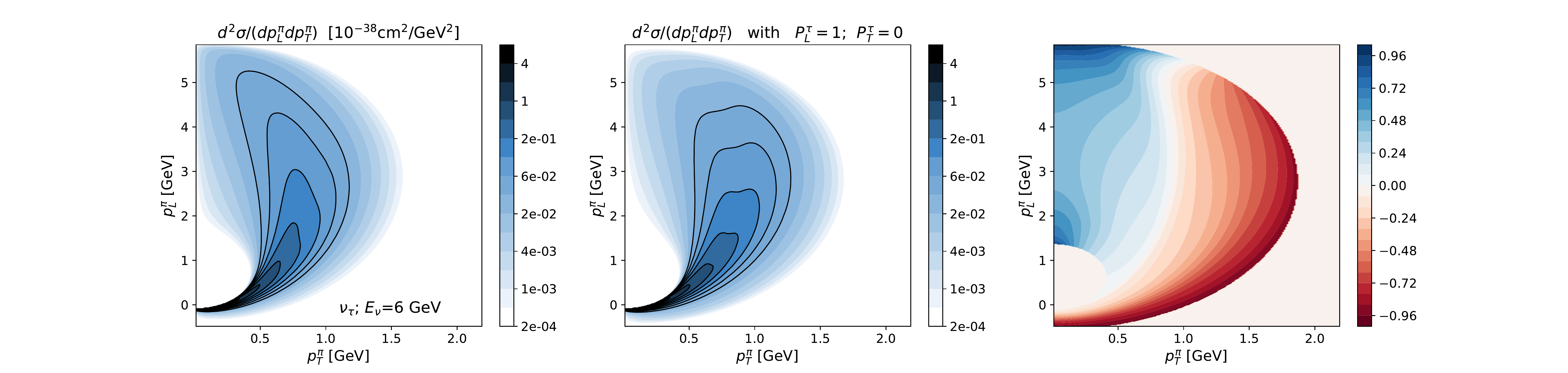}\\
\hspace*{-2cm}\includegraphics[scale=0.4]{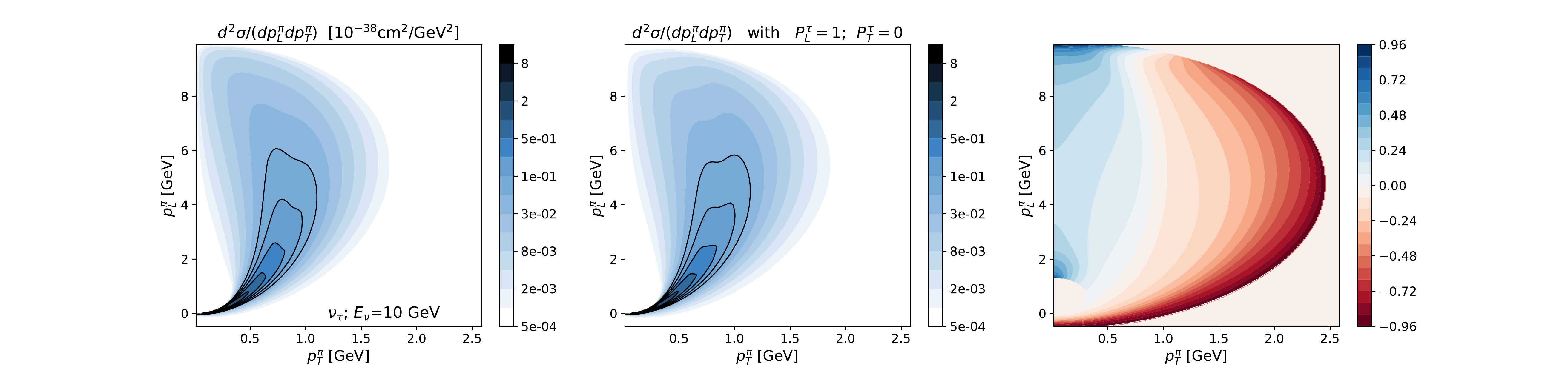}
\caption{Two-dimensional 
$d^{2}\sigma/(dp_{L\pi}dp_{T\pi})$ distributions, in units of 
$10^{-38}\,$cm$^2$/GeV$^2$, for the sequential $\nu_\tau A_Z 
\to \tau^-(\pi^- \nu_\tau) X$ process evaluated in $^{16}$O at $E_\nu=3,4,6$ 
and 10 GeV (shown from top to bottom). The left panels show the full 
calculation of Eqs.~\eqref{eq:dsdpldpt}. In the middle panels we set 
$P_L=1,P_T=0$, which corresponds in our case to pure negative-helicity taus. 
Finally, the right panels show the ratio of the difference of the two 
previous calculations over their sum.  We consider only the QE contribution 
computed using the LFG model of Ref.~\cite{Nieves:2004wx}, without including RPA and correct-energy balance corrections. } \label{fig:O16}
\end{center}
\end{figure*}

\begin{figure*}
\begin{center}
\hspace*{-2cm}\includegraphics[scale=0.4]{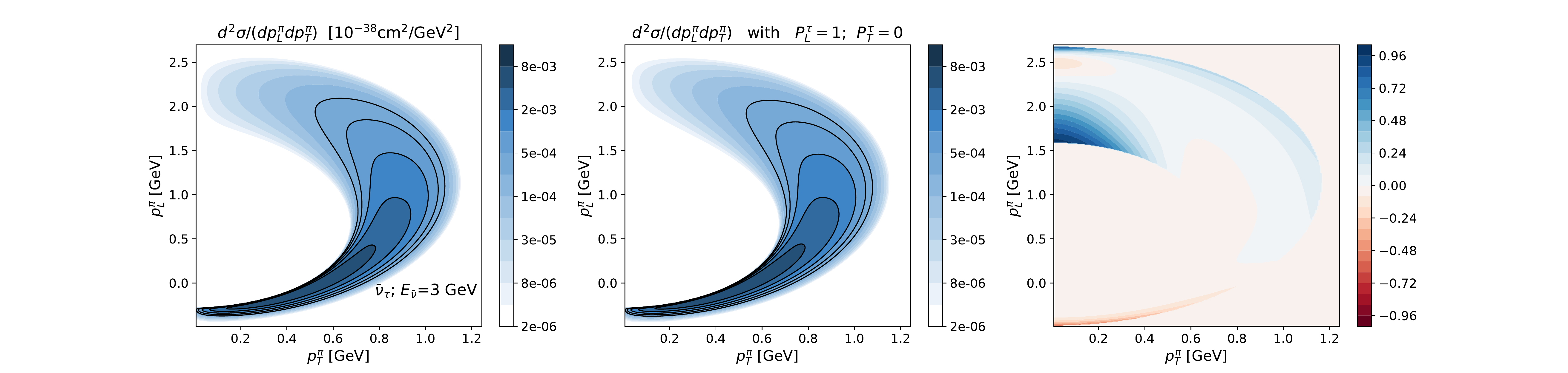}\\
\hspace*{-2cm}\includegraphics[scale=0.4]{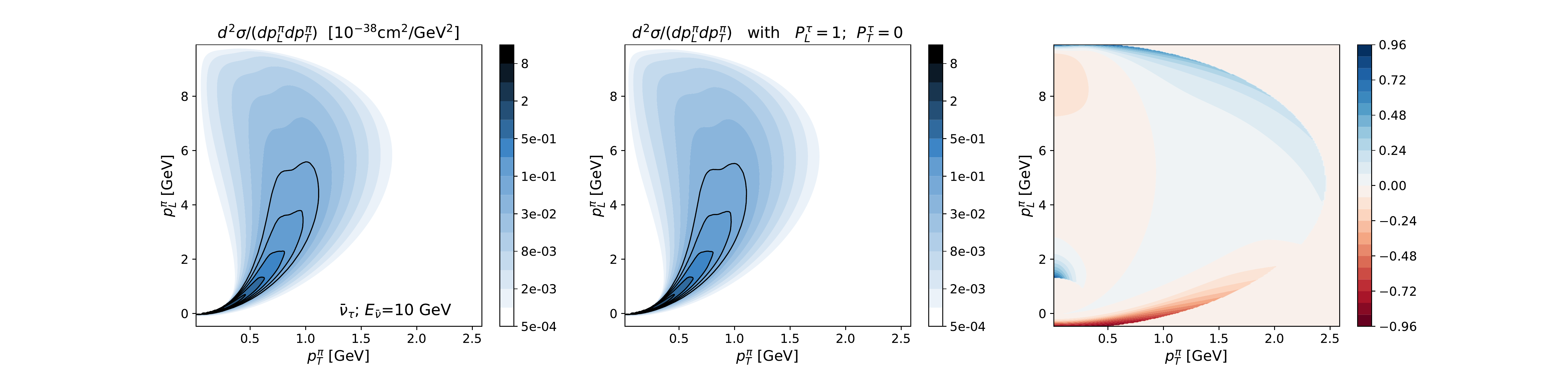}
\caption{Same as Fig.~\ref{fig:O16} but in this case for the anti-neutrino 
$\bar\nu_\tau A_Z \to \tau^+(\pi^+ \nu_\tau) X$ reaction at only $E_{\bar\nu}=3$  and 10 GeV.}\label{fig:O16-anti}
\end{center}
\end{figure*}

\begin{figure*}
\begin{center}
\hspace*{-.65cm}\includegraphics[scale=0.34]{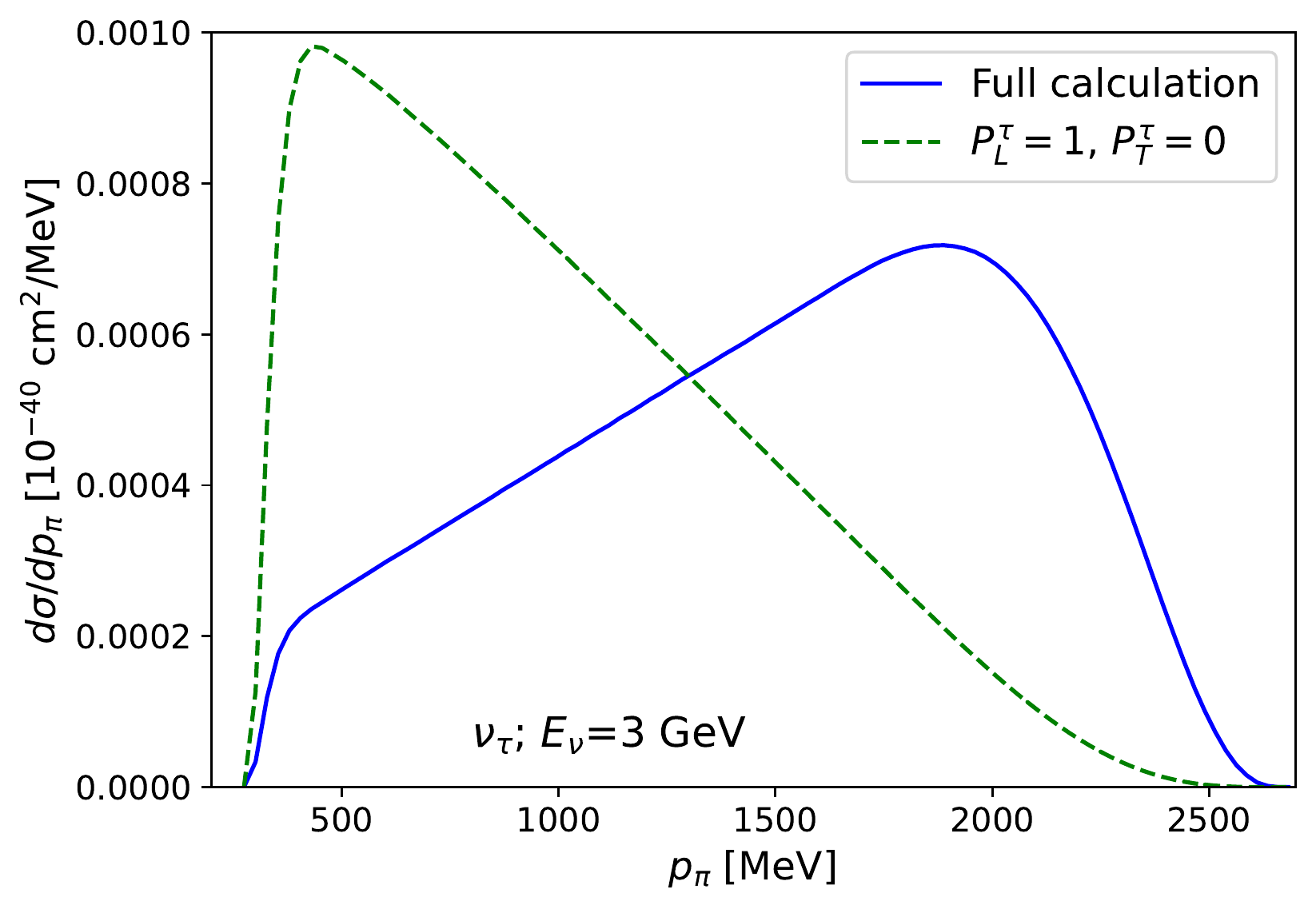}
\includegraphics[scale=0.34]{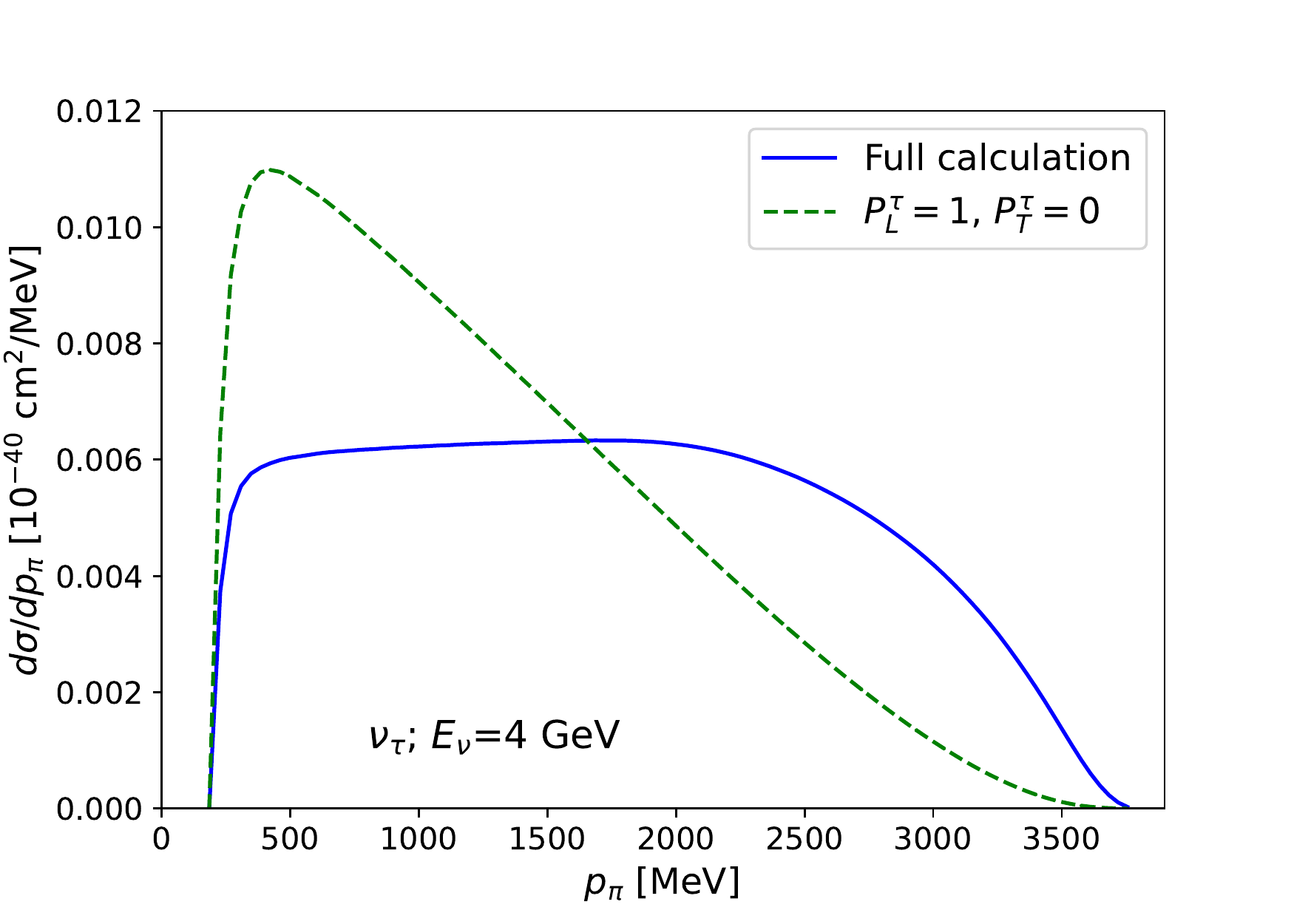}
\includegraphics[scale=0.34]{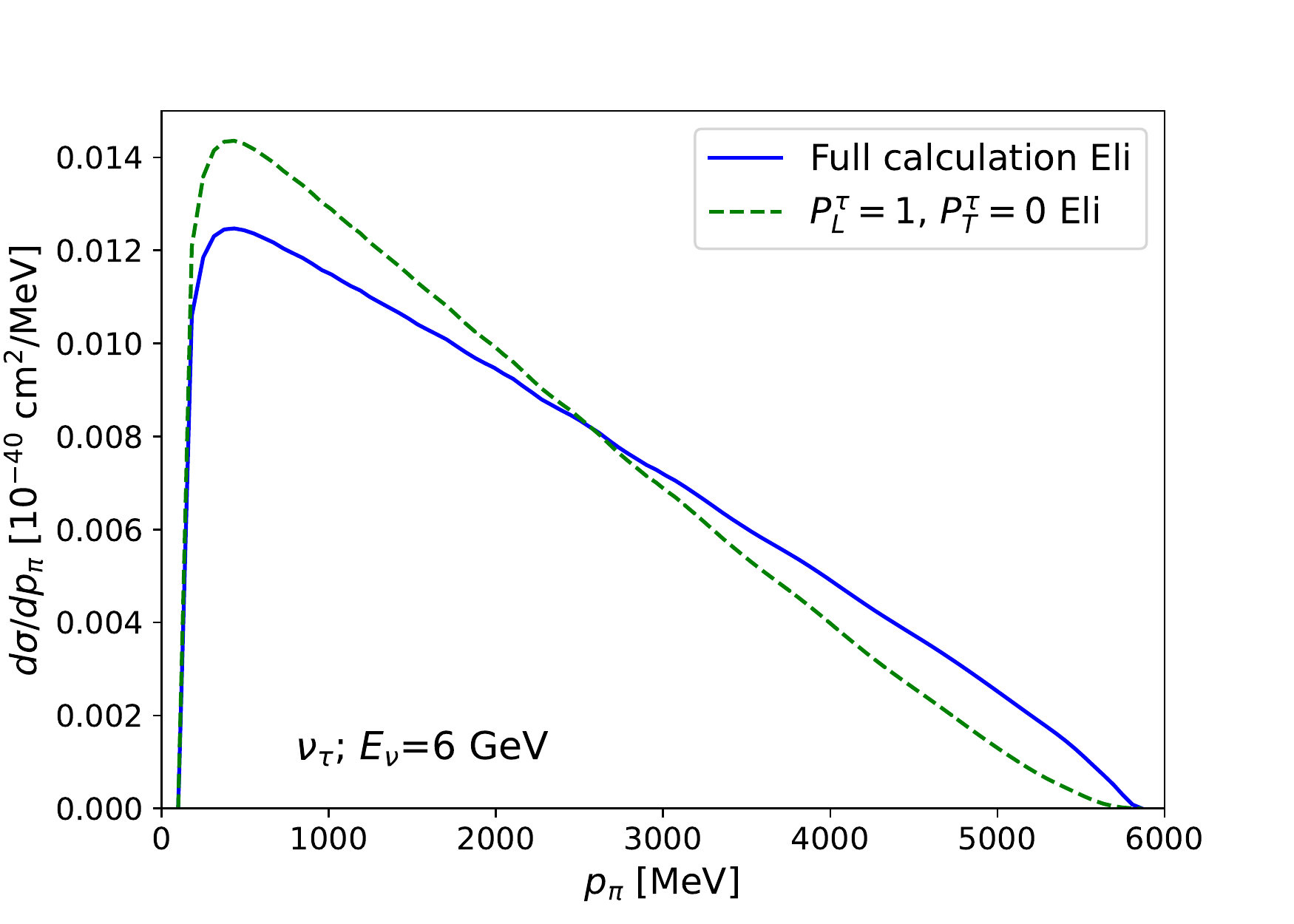} \\
\hspace*{-.65cm}\includegraphics[scale=0.34]{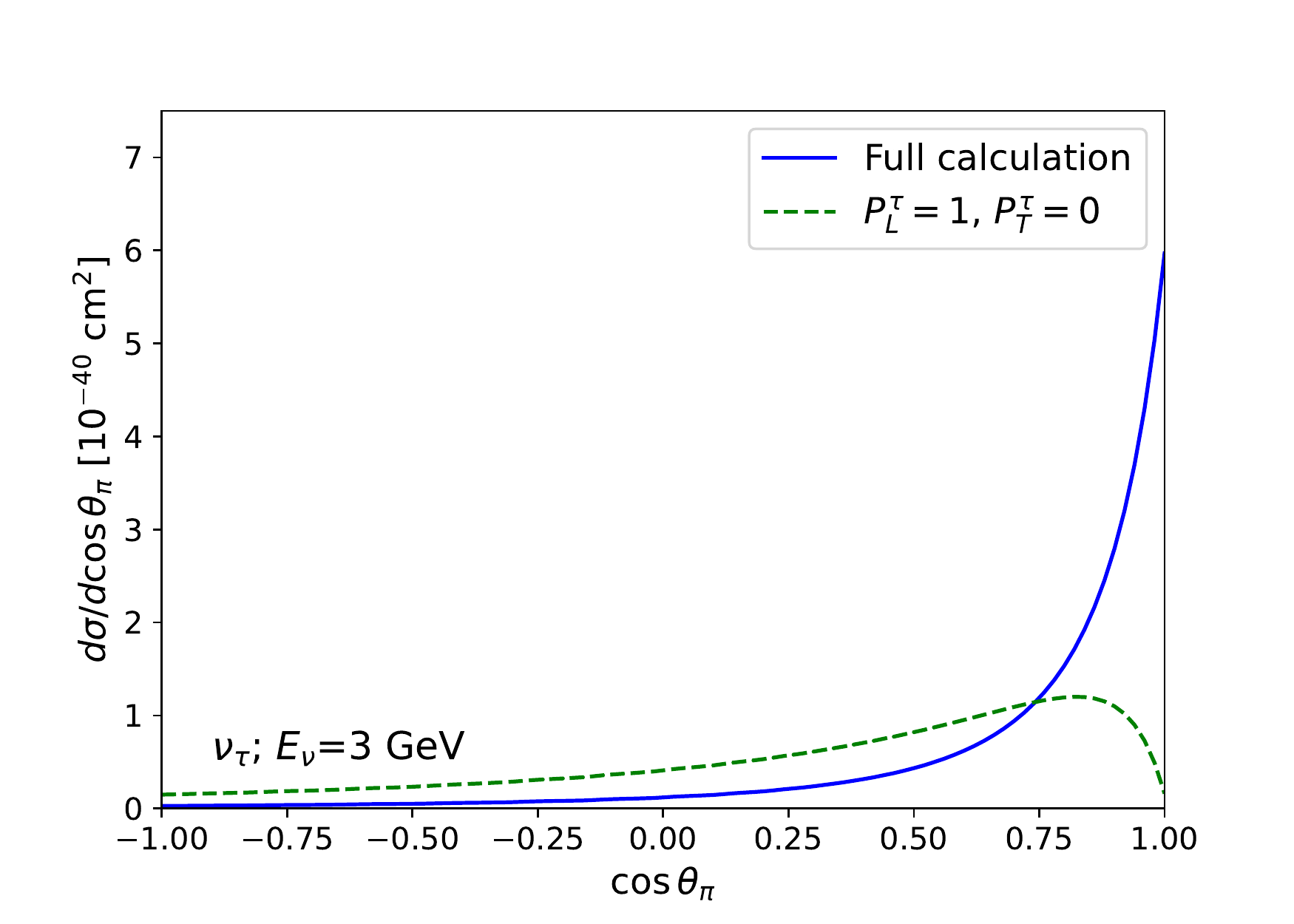}
\includegraphics[scale=0.34]{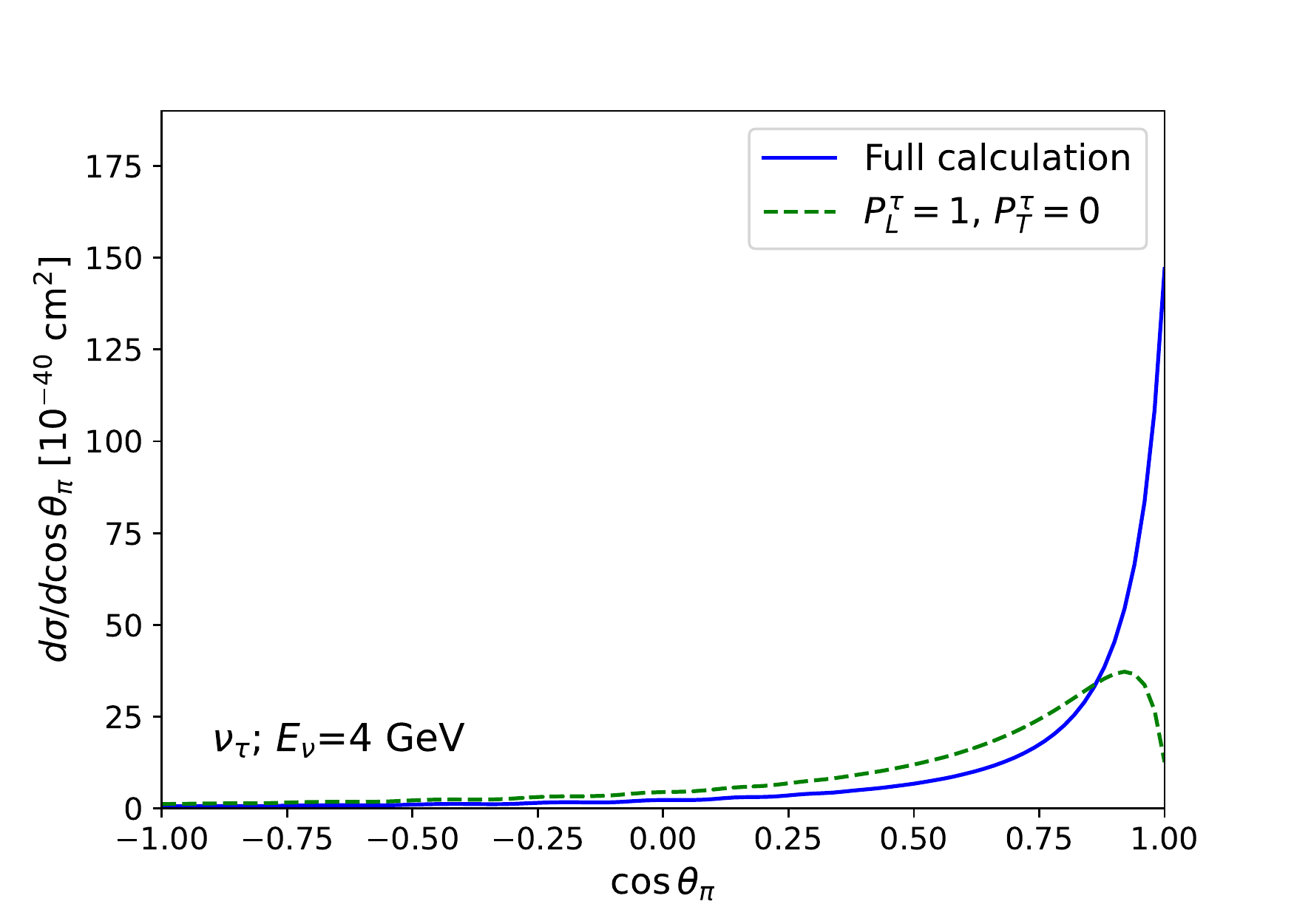}
\includegraphics[scale=0.34]{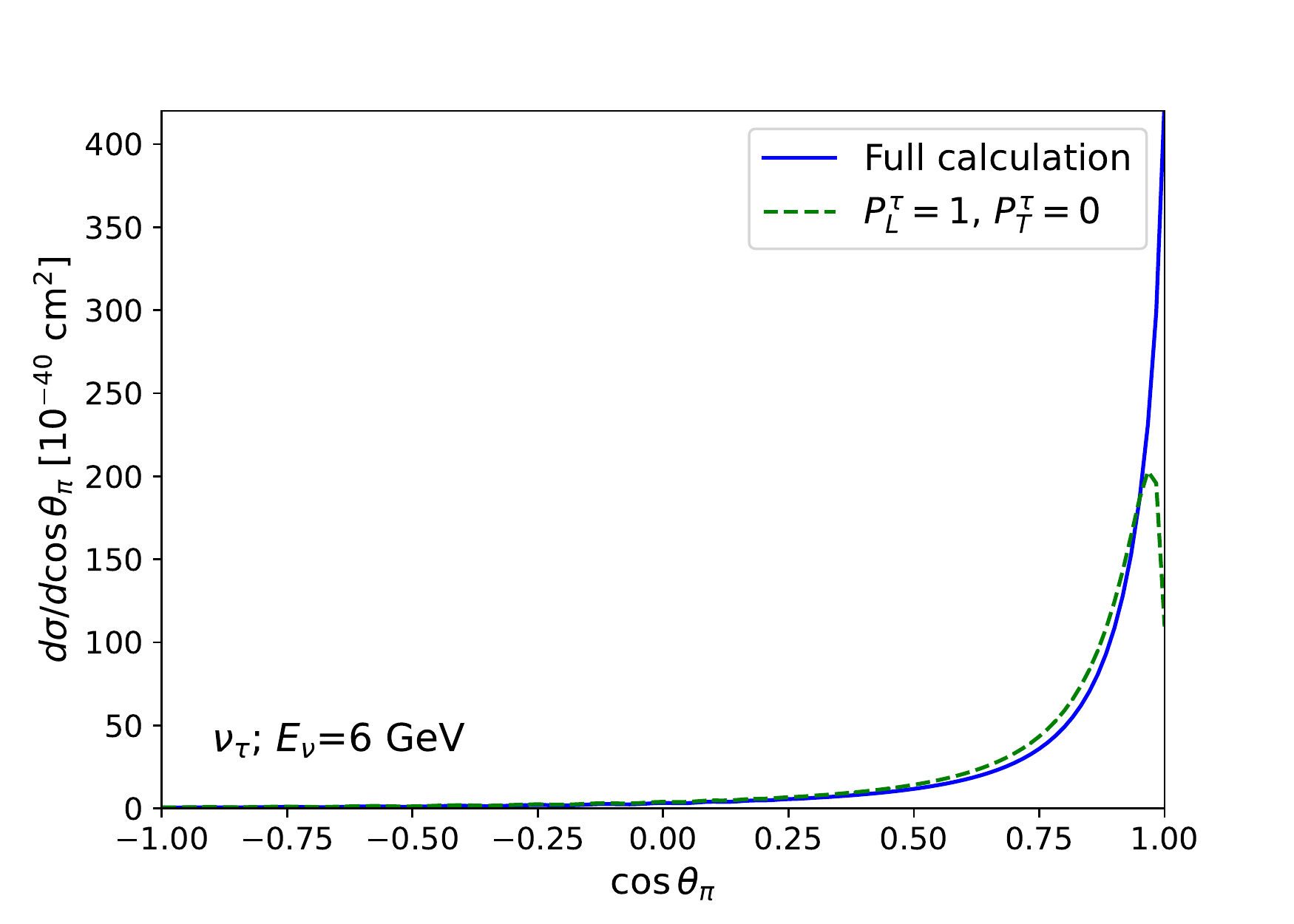} 
\caption{ $d\sigma/dp_\pi$ (upper row) and $d\sigma/d\cos\theta_\pi$ 
(lower row) differential cross section for the $\pi^-$ from the 
$\nu_\tau A_Z \to \tau^-(\pi^- \nu_\tau) X$ sequential reaction at three 
different energies  $E_{\nu}=3,4, $  and $6$~GeV. Results are shown for 
full polarization and longitudinal polarization ($P_L=1,P_T=0$), with 
the areas under both types of distributions equal, as follows from the 
discussion of Eq.~\eqref{eq:norm}. }\label{fig:1Dplots}
\end{center} 
\end{figure*}

\begin{figure*}
\begin{center}
\hspace*{-1.75cm}\includegraphics[scale=0.4]{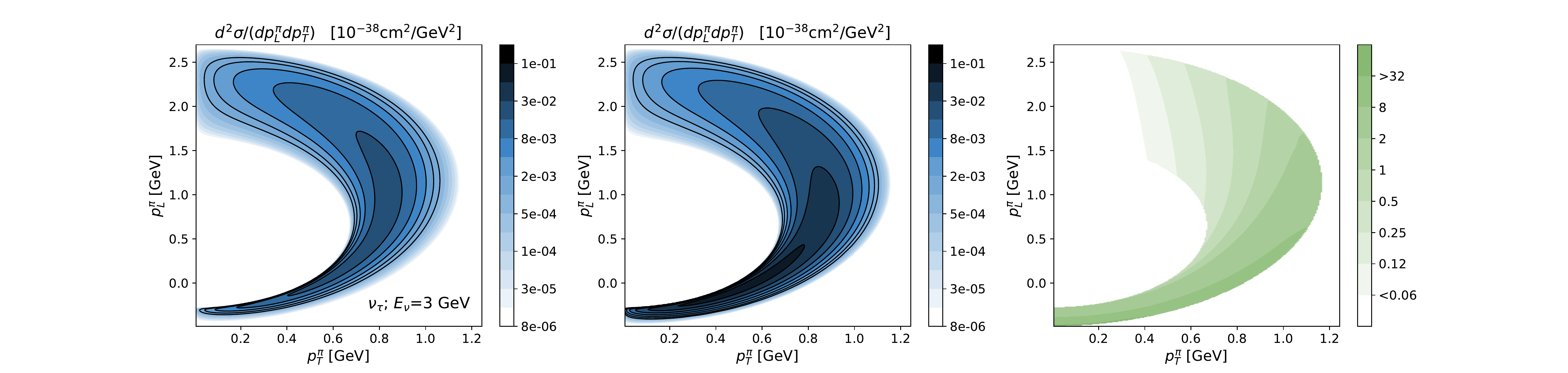}\\
\hspace*{-1.75cm}\includegraphics[scale=0.4]
{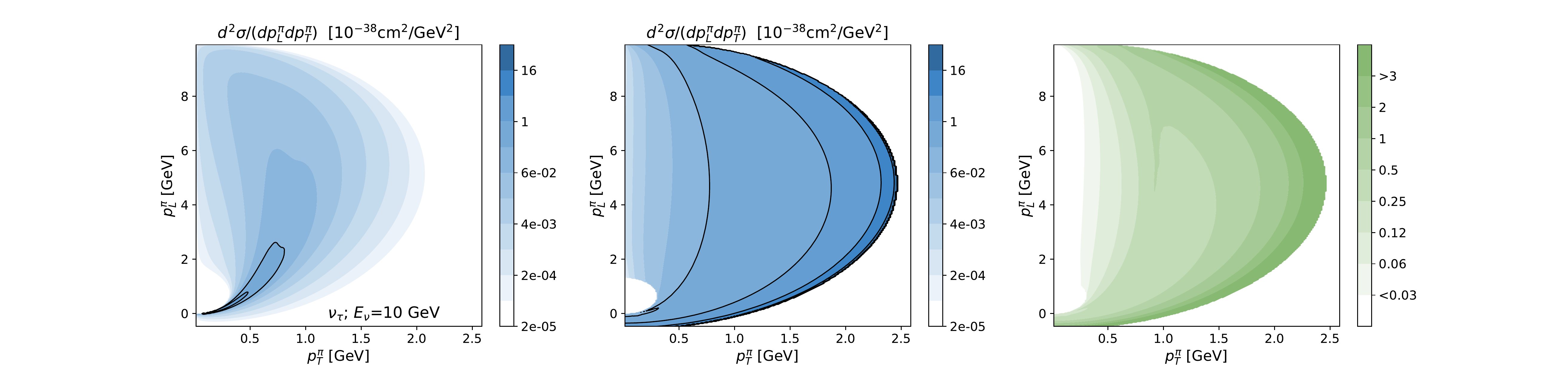}
\caption{Left panels: two-dimensional $d^{2}\sigma/(dp_{L\pi}dp_{T\pi})$ 
distribution, in units of $10^{-38}\,$cm$^2$/GeV$^2$, for the  sequential 
$\nu_\tau A_Z \to \tau^-(\pi^- \nu_\tau) X$ reaction evaluated in 
$^{16}$O at $E_{\nu}=3$ GeV (top) and 10 GeV (bottom) and obtained 
within the complete QE model. Middle panels: same as before but with 
the contributions from $W_4$ and $W_5$ switched-off. Right panels: 
the corresponding ratios as
defined in Eq.~\eqref{eq:ratio}.}\label{fig:O16noW4W5}
\end{center}
\end{figure*}

{\it We were pointed out by J. Isaacson of an error in our numerical 
calculation  in version v1 of arXiv. This affected the results shown
in Figs.~\ref{fig:O16}-\ref{fig:O16noW4W5} of that version. However, with the exception of 
a comment to  Fig.~\ref{fig:1Dplots}, that we have now corrected, 
the original text in this section  is still valid and we will not modify 
it further. The corrected figures are now provided.}\vspace{.5cm}

In this section and for illustrating purposes, we shall present results for the CCQE  $d^{2}\sigma/(dp_{L\pi}dp_{T\pi})$ distributions for the sequential $\nu_\tau A_Z \to \tau^-(\pi^- \nu_\tau) X$ and $\bar\nu_\tau A_Z \to \tau^+(\pi^+ \bar\nu_\tau) X$ processes evaluated in $^{16}$O. We focus only on the pion production, since the polarization effects for the $\rho$ particle are smaller (see Eq.~\eqref{eq:defsigmad}, $a_{d=\rho}\approx 0.4$).
The QE contribution has been computed using the LFG model of Ref.~\cite{Nieves:2004wx}, and for simplicity, we will neither include RPA-type effects, nor account for the energy balance corrections. The latter are small for the relatively moderate  neutrino energies which are considered in this work, while RPA correlations  do not appreciably change 
the gross features of the tau polarization vector. The reason is
that the polarization components are obtained as a ratio of linear combinations of nuclear structure functions and RPA changes similarly numerator and denominator~\cite{Graczyk:2004uy, Valverde:2006yi}. RPA corrections might affect the nuclear response  $F_{(\nu_\tau, \bar\nu_\tau)}$ that appears as a global factor in Eq.~\eqref{eq:defsigmad}. However, we do not expect such effects to significantly change the qualitative characteristics  of the distribution of pions from the  sequential $\tau-$decay that  will be discussed below.

The minimum tau (anti-)neutrino energy considered in this work has been 3 GeV, for which the QE CC cross section in oxygen is about a factor one-hundred  larger than the coherent pion production one, the latter estimated using the model of Ref.~\cite{Amaro:2008hd}. We have checked the coherent channel since its threshold is essentially $m_\tau$,  below that of  the QE reaction-mechanism. 

We start by showing in  Fig.~\ref{fig:O16} the results for  the   sequential $\nu_\tau A_Z \to \tau^-(\pi^- \nu_\tau) X$ process  at $E_\nu=3,4,6$ and 10 GeV (shown from top to bottom).
The left panels  show the full calculation of Eqs.~\eqref{eq:defsigmad} and \eqref{eq:dsdpldpt}, while in the middle ones we set $P^\tau_L=1$ and $P^\tau_T=0$, which corresponds in our case to pure negative-helicity taus. The latter would be similar to what is done in experimental analyses where Neutrino interaction Monte Carlo models generate the $\tau$ kinematics with models similar to the case of $\mu$ and $e$ and let the TAUOLA~\cite{Was:2004dg} package to decay the $\tau$ normally assuming fully longitudinal or fixed $\tau^{\pm}$ polarizations. In the right panels we show the ratio of the difference of the two previous calculations over their sum. As it  can be seen from the figure, at low
neutrino energies the difference between the two calculations is substantial, while it decreases at higher neutrino energies. This behaviour can be understood since high energy neutrinos produce high energy taus and the latter tend to be in a negative-helicity state due to the $(1-\gamma_5)$ part of the weak production vertex (at high energies, compared to the mass of the lepton, helicity equals chirality). However, closer to the production threshold the taus are not so energetic and the helicity is not well defined in this case. Thus, for low neutrino energies accounting for the correct tau-polarization vector is essential.

The corresponding results for the anti-neutrino $\bar\nu_\tau A_Z \to \tau^+(\pi^+ \bar\nu_\tau) X$ sequential process are shown in Fig.~\ref{fig:O16-anti}. In this case the full calculation and the one setting 
$P_L^\tau=1$ and $P_T^\tau=0$, that corresponds in our case to positive-helicity anti-taus, are very similar already at low anti-neutrino energies. This is in agreement with the findings in Ref.~\cite{Sobczyk:2019urm} and it is due to the fact that the $\bar \nu$ scattering is more forward peaked than the $\nu$ one for QE (and also for the 2p2h) mechanisms~\cite{Nieves:2013fr}. The latter effect is produced because of the different sign of  $W_3$  term, which leads to a destructive interference for the anti-neutrino case that becomes more and more effective as $|q^2|$ increases (see Eq.~D2 of Ref.~\cite{Nieves:2004wx}).

 The results discussed in this manuscript also show the relevance of the transverse polarization in future oscillation experiments such as DUNE, which expects a sizeable amount of $\nu_{\tau}$ CC interactions in the far detector \cite{Machado:2020yxl} after oscillations. To show its relevance, we present in  Fig. \ref{fig:1Dplots}, the momentum and angular distributions of the $\pi^-$ originating  from $\tau^-$ decays  for 3, 4 and 6~GeV neutrinos.
The differences between the two models discussed in this work, one with full polarization 
and the other restricted to a pure  longitudinal ($P_L=1$) lepton polarisation,  reveal t
he strong dependence of the $\pi^-$ kinematics with the transverse polarisation. 
As expected, the difference is reduced for higher momentum $\tau^-$'s. The majority of 
$\nu_{\tau}$ flux in DUNE is expected below 7~GeV\cite{Machado:2020yxl}. The case of 3~GeV 
is very conclusive. The transverse polarisation transforms a soft $\pi^-$ spectra, 
into a hard one which is typically above those produced in NC interaction. 
  
It is also interesting to evaluate the distributions in the case where $W_4=W_5=0$. As seen in Eqs.~\eqref{eq:defF},\eqref{eq:PL} and \eqref{eq:PT}, the contribution of these two structure functions is suppressed by powers of the  charged lepton  mass  and thus they can no be accessed in the corresponding reactions involving the first two lepton generations. The relevance of these structure functions is analyzed in Fig.~\ref{fig:O16noW4W5} where, for the sequential $\nu_\tau A_Z \to \tau^-(\pi^- \nu_\tau) X$ process evaluated in $^{16}$O at $E_{\nu}=3$~GeV and 10~GeV, we compare the two calculations and we further   show the ratio 
\bea
\frac{\left(\frac{d^{2}\sigma}{dp_{L\pi}dp_{T\pi}}\big|_{W_4=W_5=0}\right)-\frac{d^{2}\sigma}{dp_{L\pi}dp_{T\pi}}}{\frac{d^{2}\sigma}{dp_{L\pi}dp_{T\pi}}}
\label{eq:ratio}
\eea
 The results clearly show the significant impact of the $W_4$ and $W_5$ contributions for the case of $\tau$ production. We have checked that the effects are even more pronounced for  anti-neutrinos. They produce a general reduction in the cross section. Their effect is present for all energies analyzed although it is more relevant at lower neutrino energies.

\section{Summary}
\label{sec:summary}

We have analysed the sequential $\nu_\tau A_Z \to \tau^-(\pi^- \nu_\tau, \rho^-\nu_\tau) X$ and $\bar\nu_\tau A_Z \to \tau^+(\pi^+ \bar\nu_\tau, \rho^+ \bar\nu_\tau) X$ reactions. For the first time we give the general expression [Eq.~\eqref{eq:defsigmad}] for the outgoing hadron (pion or rho meson) energy and angular differential cross section.
Within this context, we have investigated the role of the (anti-)tau polarization vector in the analysis of these processes. We have shown that such distributions depend on the tau inclusive nuclear CC differential cross section and the longitudinal and transverse polarization observables integrated, with certain dynamical weights, over the  outgoing $\tau$ available phase space.  Since the $\tau$ momentum can not be easily reconstructed, the study of the visible kinematics of its sequential decays is the best that can be done to extract  the information on the nuclear response and reaction mechanism encoded in the polarization state of the produced tau. This information  goes beyond that obtained from the inclusive nuclear weak CC differential cross section and it can be used to further constrain nuclear models. 
 
Though all possible neutrino-nucleus reaction mechanisms contribute to the visible distribution, $d^2\sigma_d/(dE_d d\cos\theta_d)$, depending on the neutrino energy and implemented cuts, it may be possible to isolate/enhance the contribution of different (anti-)neutrino-nucleus reaction channels. Our results for the pion decay mode in oxygen at $E_\nu \le 6-10$~GeV and for the QE reaction show the relevance of considering the correct $\tau$ polarization versus some simplifications  ($P_L \sim 1$ and  $P_T \sim 0$) usually assumed in experimental neutrino oscillation analyses~\cite{Machado:2020yxl}. One exception known to us is the SuperKamiokande atmospheric $\nu_{\tau}$ analysis based on the polarization calculations given in \cite{Hagiwara:2003di}, though some approximations are still adopted. This is more significant for $\nu_\tau$-induced reactions that for the ones initiated by $\bar\nu_\tau$. In the latter case, the more forward character of the QE reaction enhances the $P_L$ component of the $\tau^+$
polarization vector, which renders the full calculation in a much better agreement with the approximate one. We have also explored the relevance of the contributions of the $W_4$ and $W_5$ nuclear structure functions. The contributions associated to these two structure functions are proportional to the charged lepton mass and, thus, they play a minor role in reactions involving the two light lepton families. However, they give a significant contribution for CC processes initiated by tau (anti-)neutrinos. Again the fact that more structure functions play a role for these sequential reactions helps in constraining nuclear models and possible reaction mechanisms.

Finally, as we have already mentioned, unless one is able to select a certain type of events by imposing a certain additional signature, all possible mechanisms contribute to $d^2\sigma_d/(dE_d d\cos\theta_d)$. Therefore, it  becomes essential to obtain the $W_{1,2,3,4,5}$ structure functions, which determine the hadron tensor $W^{\mu\nu}$, for other nuclear inclusive reaction channels (2p2h, pion production, DIS, etc.) to  correctly compute $P_L$, $P_T$ and consequently the visible distributions of the tau-decay products. 

\section*{Acknowledgements}
We warmly thank J. Isaacson for pointing out to us the
error in our numerical calculation. We thank Y. Hayato for useful discussions. J.E.S. acknowledges the support of the Humboldt Foundation through a Humboldt Research Fellowship for Postdoctoral Researchers. This research has been supported  by the Spanish Ministerio de Ciencia e Innovaci\'on (MICINN)
and the European Regional Development Fund (ERDF) under contracts PID2020-112777GB-I00 and PID2019-105439G-C22, 
the EU STRONG-2020 project under the program H2020-INFRAIA-2018-1, 
grant agreement no. 824093 and by  Generalitat Valenciana under contract PROMETEO/2020/023, the Deutsche
Forschungsgemeinschaft (DFG)
through the Cluster of Excellence ``Precision Physics, Fundamental
Interactions, and Structure of Matter" (PRISMA$^+$ EXC 2118/1) funded by the
DFG within the German Excellence Strategy (Project ID 39083149) and the Swiss
National Foundation under the grant 200021\_85012.

\bibliography{neutrinos}

\end{document}